\documentclass[preprint,showpacs,showkeys,amsmath,amssymb,epsfig]{revtex4}
\usepackage{epsfig}

\newcommand{\adoa}[1]{\frac{\dot a_{#1}}{a_{#1}}}
\newcommand{\addoa}[1]{\frac{\ddot a_{#1}}{a_{#1}}}

\begin{document}



\title[]{Cosmological Implication of Antisymmetric Tensor Field
on D-brane}
\author{Inyong \surname{Cho}}
\email{iycho@phya.snu.ac.kr}
\affiliation{Center for Theoretical Physics, School of Physics,
Seoul National University, Seoul 151-747, Korea}
\author{Eung Jin \surname{Chun}}
\email{ejchun@kias.re.kr} \affiliation{Korea Institute for
Advanced Study, 207-43 Cheongryangri-2-dong Dongdaemun-gu, Seoul
130-722, Korea}
\author{Hang Bae \surname{Kim}}
\email{hbkim@hanyang.ac.kr}
\affiliation{Particle and Astrophysics Research Center, Department of
Physics,
Hanyang University, Seoul 133-791, Korea}
\author{Yoonbai \surname{Kim}}
\email{yoonbai@skku.edu}
\affiliation{BK21 Physics Research Division and Institute of Basic Science,
Sungkyunkwan University, Suwon 440-746, Korea}



\begin{abstract}
We discuss the cosmological implications of an antisymmetric tensor field
when it experiences a kind of ``Higgs mechanism,'' including
nonlocal interaction terms, on D-branes. Even when a huge
magnetic-condensation-inducing anisotropy is assumed in the early
universe, the expansion of the universe leads to an isotropic
$B$-matter-dominated universe.
\end{abstract}


\maketitle

\section{Introduction}
If we are interested in the very early universe near or slightly
below the Planck scale, the cosmological model driven by string
theory should be taken into account. The first signal of string
cosmology is given by dilaton gravity instead of Einstein
gravity~\cite{Pol}. In string theory, another indispensable
bosonic degree is the antisymmetric tensor field of rank
two~\cite{Kalb:1974yc,Cremmer:1973mg}. When the ten-dimensional
bulk is compactified to a (1+3)-dimensional spacetime, the cosmological
effect of the massless antisymmetric tensor field has been taken
into account.
The homogeneous, but anisotropic, Bianchi-type
universe is supported due to the antisymmetric tensor
field~\cite{Goldwirth:1993ha}. Once the anisotropy is generated,
its effect can usually survive in the present universe, despite
the sufficient expansion of the universe. The observed data of
cosmic microwave background radiation (CMBR) threatens the viability
of this string-driven cosmological model with the antisymmetric
tensor field.

The development of D-branes and related topics for the last ten years
has opened another possibility in cosmology that our universe may be identified with
a D3-brane, or parts of higher-dimensional D-branes in a nine-dimensional
spatial bulk. A clear distinction between string
cosmology~\cite{copeland:1995} in the bulk without D-brane
and that on the D-brane appears through the mass
generation of the antisymmetric tensor field on the
D-brane~\cite{Witten:1995im}.
In the context of Einstein gravity, the cosmological effect of this massive
antisymmetric field was studied in detail~\cite{Chun:2005ee}.
Specifically, the anisotropy
induced by condensation of the antisymmetric tensor field is diluted
during the expansion of the
universe~\cite{Kim:2004zq}. If the expansion is sufficient, the resultant
present universe becomes isotropic and is free from the constraint of the CMBR.

In this paper, we will investigate the effect of the massive interacting
antisymmetric tensor field. In Section II, we introduce the model of our
interest, including graviton, dilaton, and antisymmetric tensor field
in both the string frame and the Einstein frame.
In Section III, we consider only
a single magnetic component of the homogeneous antisymmetric tensor field
and study its oscillation without damping in a flat spacetime.
In Section IV, we consider another flat system of the antisymmetric tensor
field with a dilaton. For the case of negative cosmological constant,
the stabilization of the dilaton is achieved for some initial values.
In Section V, the cosmological evolution of the antisymmetric tensor field
is studied without and with a D3-brane. In the case of a D-brane universe,
solutions of the isotropic $B$-matter-dominated universe are obtained.
We conclude in Section VI with a summary and discussion.

\section{String Effective Theory in a D3-brane Universe}
When a ten-dimensional string theory is compactified to a
four-dimensional theory, the spectrum of particles depends on a
specific pattern of the compactification, and the presence of
(D-)branes generates various models, including a model that
resembles the standard model. In order to understand the cosmological
evolution, we take into account closed strings containing the
graviton. The universal nature of cosmology based on
four-dimensional effective field theory requires bosonic particles
at the lowest energy level be produced irrespective of the
compactification pattern. Then, we mainly examine  the cosmological
effect due to the presence of a D3-brane whose world-volume is
identified with our spacetime.

Bosonic degrees from the closed strings of string theory include
the graviton $\tilde{g}_{\mu\nu}$, the dilaton $\Phi$, and the antisymmetric
tensor field $B_{\mu\nu}$. On the D3-brane forming our
universe, the sum of the antisymmetric tensor field $B_{\mu\nu}$ and
the field strength tensor $F_{\mu\nu}$ of a U(1) gauge field $A_{\mu}$
defines a gauge invariant ${\cal
B}_{\mu\nu}=B_{\mu\nu}+2\pi\alpha' F_{\mu\nu}$~\cite{Pol}. Their
four-dimensional effective action in the string frame is given by
the sum of the bulk action and the Dirac-Born-Infeld (DBI)-type brane
action:
\begin{eqnarray}\label{act1a}
S_{4}&=& \frac{1}{2\kappa_4^2} \int d^4\tilde x \sqrt{-\tilde g}
\left[ e^{-2\Phi}\left(\tilde
R-2\Lambda
\right.\right.
\nonumber\\
{} &+& \left. 4\tilde\nabla_\mu\Phi\tilde\nabla^\mu\Phi
- \frac{1}{12} H_{\mu\nu\rho} H^{\mu\nu\rho}\right)
\vphantom{\sqrt{\frac12}}
\\
{} &-& \left.{m_B^2}e^{-\Phi} \sqrt{1 + \frac12{\cal B}_{\mu\nu}{\cal
B}^{\mu\nu} - \frac{1}{16}\left({\cal B}^*_{\mu\nu}{\cal
B}^{\mu\nu} \right)^2}\ \right],\nonumber
\end{eqnarray}
where ${\cal B}^{\ast}_{\mu\nu}= \sqrt{-\tilde
g}\epsilon_{\mu\nu\alpha\beta} {\cal B}^{\alpha\beta}/2$ with
$\epsilon_{0123} = 1$ and ${H}_{\mu\nu\rho}=\partial_{[\mu} {\cal
B}_{\nu\rho ]}$. $m_B$ is a parameter defined by
$m_B^2=2\kappa_4^2{\cal T}_3$ where ${\cal T}_3$ is the tension of the
D3-brane.
The action in Eq.~(\ref{act1a}) leads to the following equations of
motion for the antisymmetric tensor field, the dilaton, and the
graviton, respectively:
\begin{eqnarray}
\label{B-eq0}
\tilde{\nabla}^\lambda {H}_{\lambda\mu\nu}
&-& 2{H}_{\lambda\mu\nu} \tilde{\nabla}^\lambda\Phi \\
{}&-& m_B^2e^{\Phi}\frac{{\cal B}_{\mu\nu}-\frac14{\cal B}^*_{\mu\nu}
\left({\cal B}{\cal B}^*\right)}{\sqrt{1+\frac12{\cal B}^2
-\frac{1}{16}\left({\cal B}{\cal B}^*\right)^2}} = 0,\nonumber
\end{eqnarray}
\begin{eqnarray}
\label{D-eq0}
4\tilde{\nabla}^2\Phi-4\left(\tilde{\nabla}\Phi\right)^2 &=&
-\tilde{R}+2\Lambda+\frac{1}{12}{H}^2 \\
{}&+&\frac12m_B^2e^{\Phi}
\sqrt{1+\frac12{\cal B}^2-\frac{1}{16}\left({\cal B} {\cal
B}^*\right)^2}\, ,\nonumber
\end{eqnarray}
\begin{eqnarray}
\label{E-eq0} \tilde{G}_{\mu\nu} = -\tilde{g}_{\mu\nu}\Lambda +
\kappa_4^2 \tilde{T}_{\mu\nu} ,
\end{eqnarray}
where the energy-momentum tensor is
\begin{eqnarray}
\tilde{T}_{\mu\nu}&=& \tilde{T}^{\Phi}_{\mu\nu}+\tilde{T}^{{\cal
B}}_{\mu\nu},
\label{em}\\
\kappa_4^2 \tilde{T}^{\Phi}_{\mu\nu}&=&
2\tilde{g}_{\mu\nu}\left[\tilde{\nabla}^2\Phi
-\left(\tilde{\nabla}\Phi\right)^2\right]-2\tilde{\nabla}_\mu
\tilde{\nabla}_\nu\Phi ,\label{TPhi}\\
\kappa_4^2 \tilde{T}^{{\cal B}}_{\mu\nu}&=&\frac{1}{12}
\left(3{H}_{\mu\lambda\rho}{H}_\nu^{\;\lambda\rho}
-\frac12\tilde{g}_{\mu\nu}{H}^2\right) \label{TBm}\\
{}&+&\frac12m_B^2e^\Phi
\frac{-\tilde{g}_{\mu\nu}-\frac12\tilde{g}_{\mu\nu} {\cal B}^2
+{\cal B}_{\mu\lambda}{\cal B}_\nu^{\;\lambda}}
{\sqrt{1+\frac12{\cal B}^2-\frac{1}{16}\left({\cal B} {\cal
B}^*\right)^2}}.
\nonumber
\end{eqnarray}

Since we are interested in the time evolution, specifically the expansion
of the universe, a natural scheme may require the reproduction of
Einstein gravity at late times. In the classical level, the model of
our interest in the string frame of Eq.~(\ref{act1a}) does not exhibit
stabilization of the dilaton $\Phi$, so one option is to study
this topic in the Einstein frame obtained by a coordinate
transformation from the string frame,
\begin{equation}
g_{\mu\nu} = e^{-2\Phi}\tilde{g}_{\mu\nu} .
\end{equation}
Then, from Eq.~(\ref{act1a}), we obtain the action
\begin{eqnarray}\label{action-E}
S_{\rm E}&=& \frac{1}{2\kappa_4^2} \int d^4x \sqrt{-g} \left[ R
-2\Lambda e^{2\Phi}
\right.\nonumber\\
{}&-& 2(\nabla\Phi)^2
-\frac{1}{12}e^{-4\Phi}H^2 \\
\vphantom{\sqrt{\frac12}}
{}&-& \left. {m_B^2}e^{3\Phi}\sqrt{1 + \frac12e^{-4\Phi}{\cal B}^2 -
\frac{1}{16}e^{-8\Phi}\left({\cal B}^*{\cal B}\right)^2} \right].\nonumber
\end{eqnarray}

In the Einstein frame, we read again the equations of motion for
the antisymmetric field ${\cal B}_{\mu\nu}$, the dilaton $\Phi$,
and the graviton $g_{\mu\nu}$:
\begin{eqnarray}\label{Be}
\nabla_\lambda H_{\lambda\mu\nu}
&-&4H_{\lambda\mu\nu}\nabla^\lambda\Phi \\
{}&-& m_B^2e^{3\Phi}\frac{{\cal
B}_{\mu\nu}-\frac14e^{-4\Phi}{\cal B}^*_{\mu\nu} \left({\cal
B}{\cal B}^*\right)}{\sqrt{1+\frac12e^{-4\Phi}{\cal B}^2
-\frac{1}{16}e^{-8\Phi}\left({\cal B}{\cal B}^*\right)^2}} = 0,\nonumber
\end{eqnarray}
\begin{equation}\label{De}
\nabla^2\Phi = \frac{\partial V(\Phi)}{\partial\Phi},
\end{equation}
where the potential is
\begin{eqnarray}
V(\Phi) &=& {1\over 4}\left[ 2\Lambda e^{2\Phi} +
\frac{1}{12}e^{-4\Phi}H^2 \vphantom{\sqrt{\frac12}}
\right.\\
{}&+& \left. m_B^2e^{3\Phi}\sqrt{1 + \frac12e^{-4\Phi}{\cal B}^2 -
\frac{1}{16}e^{-8\Phi}\left({\cal B}^*{\cal B}\right)^2}\right],\nonumber
\end{eqnarray}
and
\begin{eqnarray}\label{Ee}
G_{\mu\nu} = -g_{\mu\nu}\Lambda e^{2\Phi}+ \kappa_4^2 T_{\mu\nu} ,
\end{eqnarray}
where the energy-momentum tensor is
\begin{eqnarray}
T_{\mu\nu}&=& T^{\Phi}_{\mu\nu}+T^{{\cal B}}_{\mu\nu},
\label{emE}\\
\kappa_4^2
{T}^{\Phi}_{\mu\nu}&=&2\nabla_\mu\Phi\nabla_\nu\Phi-g_{\mu\nu}(\nabla\Phi)^2,
\label{TPE}\\
\kappa_4^2 {T}^{{\cal B}}_{\mu\nu}&=&\frac{1}{12}e^{-4\Phi}
\left(3H_{\mu\lambda\rho}H_\nu^{\;\lambda\rho}
-\frac12g_{\mu\nu}H^2\right) \label{TBE}\\
{}&+&\frac12m_B^2e^{3\Phi}
\frac{-g_{\mu\nu}-\frac12g_{\mu\nu}e^{-4\Phi}{\cal B}^2
+e^{-8\Phi}{\cal B}_{\mu\lambda}{\cal B}_\nu^{\;\lambda}}
{\sqrt{1+\frac12e^{-4\Phi}{\cal B}^2
-\frac{1}{16}e^{-8\Phi}\left({\cal B}{\cal B}^*\right)^2}}.\nonumber
\end{eqnarray}
Note that conservation of energy-momentum is consistent with
the equations of motion in each frame,
(\ref{B-eq0})--(\ref{D-eq0}) or (\ref{Be})--(\ref{Ee}),
respectively.

A few topics will be discussed at the quantum level in
the subsequent sections, but most of the cosmological applications
will be made at the classical level. For the antisymmetric tensor
field of second rank ${\cal B}_{\mu\nu}$, we employ the
terminology of the {\it electric} components for ${\cal B}_{i0}\equiv
({\bf E})^{i}$ and of the {\it magnetic} components for ${\cal
B}_{ij}\equiv 2\epsilon_{0ijk}({\bf B})^{k}$ in what follows. In
the subsequent sections, we only consider the homogeneous magnetic
component, ${\bf E}=0$ and ${\bf B}={\bf B}(t)\ne 0$, with a fixed
direction
\begin{equation}\label{mag}
{\bf B}=B(t){\bf {\hat k}},
\end{equation}
and its time evolution in a flat spacetime.

\section{Homogeneous $B$-matter without Dilaton in Flat Spacetime}
Since our main goal is to understand the effect of the antisymmetric tensor
field in the early universe near the Planck scale, we begin this section with
studying it in flat spacetime
with a stabilized dilaton.
Substituting the magnetic field ansatz in Eq.~(\ref{mag}) with the flat metric
$g_{\mu\nu}=\eta_{\mu\nu}$ and the stabilized dilaton $\Phi=0$
into the Euler-Lagrange equation of the antisymmetric
tensor field, Eq.~(\ref{B-eq0}) in the string frame or Eq.~(\ref{Be})
in the Einstein frame, we obtain
the single dynamical equation of the $B$ field
\begin{equation}\label{seqB}
\partial_{0}^{2}B= -m_{B}^{2}
\frac{B}{\sqrt{1+B^{2}}}.
\end{equation}
Integration of Eq.~(\ref{seqB}) gives
\begin{equation}\label{foeq}
{\cal E}=\frac{1}{2}{\dot B}^{2}+V_{{\rm eff}}(B),
\end{equation}
where ${\cal E}$ is an integration constant, $V_{{\rm eff}}(B)
=\sqrt{1+B^{2}}\,$, and the overdot represents  differentiation with respect to
the rescaled time variable ${\tilde t}=m_{B}t$.
Equation (\ref{foeq}) can be rewritten as an integral equation
\begin{equation}\label{Bs2}
{\tilde t}-{\tilde t}_{0}=\pm\int_{B_{0}}^{B}
\frac{dB}{\sqrt{2({\cal E}-\sqrt{1+B^{2}}\, )}}.
\end{equation}
The only pattern of nontrivial solutions is an oscillating one with the
amplitude $\sqrt{{\cal E}^{2}-1}$ for ${\cal E}>1$. This
oscillating solution with a fixed amplitude is natural at the classical
level in a flat spacetime when a homogeneous condensation of the
magnetic field is given by an initial condition. In expanding
universes, the solution is expected to change to that with an
oscillation and a damping due to the growing of a spatial scale
factor, which means a dilution of $B$-matter in the expanding D3-brane
universe.

\begin{center}
\section{Homogeneous $B$-matter with Linear Dilaton in Flat Spacetime}
\end{center}
If we turn on the linear dilaton $\Phi$ in flat spacetime, the equations of
motion for the antisymmetric tensor field $B_{\mu\nu}$ and the dilaton
$\Phi$ become different in the string frame, Eqs.~(\ref{B-eq0})--(\ref{D-eq0}),
and the Einstein frame, Eqs.~(\ref{Be})--(\ref{De}). In this section, we again
discuss the evolution of the magnetic field with a fixed direction for both
frames.

In the string frame, plugging the ansatz for the magnetic field,
Eq.~(\ref{mag}), and the homogeneous dilaton
\begin{equation}\label{di}
\Phi=\Phi(t)
\end{equation}
into the equations of motion,  Eqs.~(\ref{B-eq0})--(\ref{D-eq0}),
gives
\begin{eqnarray}
\ddot{B}-2\dot{B}\dot\Phi
&=& -e^{\Phi}
\frac{B}{\sqrt{1+B^{2}}},
\label{m1}\\
\ddot\Phi-\dot\Phi^{2}-
\frac{1}{8}\dot{B}^{2}&=&
-\frac{1}{8}e^{\Phi}
\sqrt{1+{B}^{2}}-\frac{\tilde\Lambda}{2},
\label{m2}
\end{eqnarray}
where we rescaled $\tilde\Lambda \equiv \Lambda/m_B^2$.
Plugging the ansatz for the magnetic field,
Eq.~(\ref{mag}), and the homogeneous dilaton, Eq.~(\ref{di}),
into the equations of motion, Eqs.~(\ref{Be})--(\ref{De}), in the Einstein frame,
we have
\begin{eqnarray}
\ddot{B}-4\dot{B}\dot\Phi
&=&-e^{3\Phi}
\frac{B}{\sqrt{1+B^{2}e^{-4\Phi}}},
\label{n1}\\
\ddot\Phi+\frac{1}{2}e^{-4\Phi}\dot{B}^{2}
&=&-\frac{1}{4}e^{3\Phi}\sqrt{1+B^{2}e^{-4\Phi}}\nonumber\\
{}&-&\frac{1}{2}\frac{e^{3\Phi}}{\sqrt{1+B^{2}e^{-4\Phi}}}-\tilde\Lambda e^{2\Phi}.
\label{n2}
\end{eqnarray}

We solve the field equations for $B$ and $\Phi$ in both frames
with different values of $\tilde\Lambda$. In the string frame, for
$\tilde\Lambda =0$, 
an initial oscillation in $B$ discussed in the previous section is
observed. Later, $B$ settles down to a constant. The dilaton
$\Phi$ decreases to negative infinity.

\begin{figure}[ht]
\centerline{\epsfig{figure=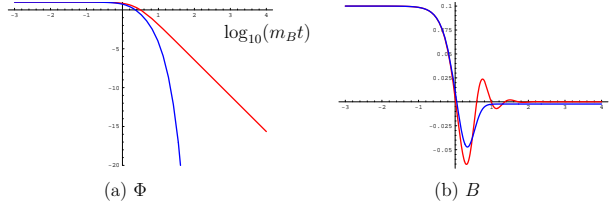,height=27mm}}
\vspace{0.1in} \caption{$\Phi$ and $B$ in the string frame for
$\tilde\Lambda =0$ and $\tilde\Lambda = 1/2$. The initial conditions are
$\Phi(0)=1$ and $B(0)=0.1$. For $\tilde\Lambda >0$, $\Phi$ drops very
rapidly, and the oscillation in $B$ is suppressed.} \label{fig1}
\end{figure}

For $\tilde\Lambda >0$, the $\tilde\Lambda$-term in the $\Phi$-equation,
Eq.~(\ref{m2}), effectively provides a potential term which is linear
in $\Phi$. Therefore, the dilaton is pushed to a negative value
more rapidly, and the oscillation in $B$ is suppressed. The
numerical solutions are plotted in Fig.~\ref{fig1}.

In the Einstein frame, in order to analyze the dilaton behavior,
let us assume $B=0$. Then, from Eq.~(\ref{n2}), we see that the system
is frictionless (no $\dot\Phi$-term) and that the dilaton field
$\Phi$ has an effective  potential
\begin{equation}
V_e = {1\over 4}e^{2\Phi} (e^\Phi +2\tilde\Lambda ).
\end{equation}
For $\tilde\Lambda \geq 0$, $V_e$ is a monotonically increasing function
of $\Phi$, which pushes $\Phi$ to  negative infinity. Once $\Phi$
achieves sufficiently large negative values, the $\tilde\Lambda$-term in
the dilaton  equation, Eq.~(\ref{n2}), in the Einstein frame becomes
negligible.

\begin{figure}[ht]
\centerline{\epsfig{figure=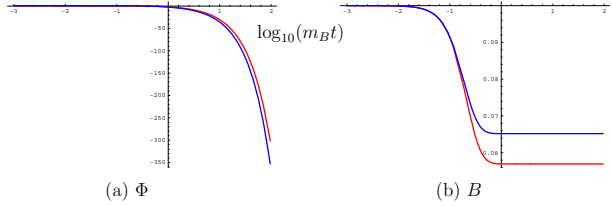,height=27mm}} \vspace{0.1in}
\caption{$\Phi$ and $B$ in the Einstein frame for $\tilde\Lambda =0$ and
$\tilde\Lambda = 1/2$. The initial conditions are $\Phi(0)=1$ and
$B(0)=0.1$. The configurations are very similar regardless of
$\tilde\Lambda$.} \label{fig2}
\end{figure}

\begin{figure}[t]
\centerline{\epsfig{figure=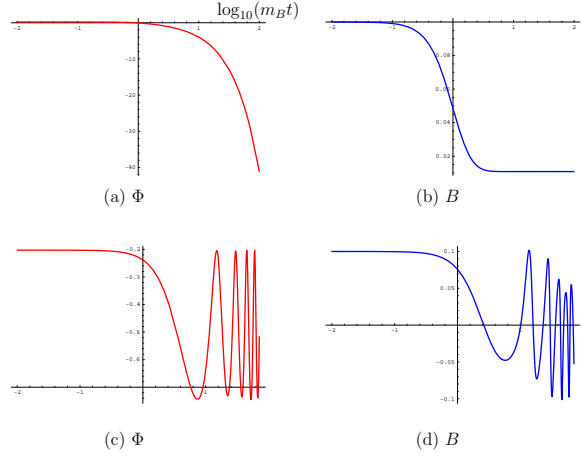,height=60mm}}
\vspace{0.1in} \caption{$\Phi$ and $B$ in the Einstein frame for
$\tilde\Lambda =-1/2$ and $B(0)=0.1$. For large $\Phi(0)=0.202733$
(upper panel), the configurations are very similar to those for
$\tilde\Lambda \geq 0$. For small $\Phi(0)=-0.202733$ (lower panel), the
dilaton stabilization is observed.} \label{fig3}
\end{figure}

For $\tilde\Lambda <0$, $V_e$ possesses a global minimum at $\Phi_s =
\ln(-4\tilde\Lambda/3)$ and approaches zero from below as $t \to
-\infty$. If the initial $\Phi(0) \equiv \Phi_0$ is imposed such
that $V(\Phi_0) \geq 0$, the dilaton $\Phi$ rolls down to the
minimum and is pushed to negative infinity eventually. However, if
$\Phi_0$ is taken such that $V(\Phi_0) < 0$, $\Phi$ will oscillate
about $\Phi_s$, and the dilaton is stabilized. Even after we turn
on the $B$-field, the story is not altered very much. We observe
this with the $B$-field included, from the numerical solutions
plotted in Fig.~\ref{fig2} for $\tilde\Lambda \geq 0$ and
Fig.~\ref{fig3} for $\tilde\Lambda < 0$.

\section{Homogeneous Universe Dominated by $B$-matter}

The main topic of our interest is cosmological implications of
large-scale fluxes of the antisymmetric tensor field which might have
existed in the early universe of a string theory scale.
When the early universe is dominated by the massive antisymmetric
tensor field in appropriate initial configurations, its cosmological
evolution will be traced.

We take into account the same homogeneous magnetic ansatz as in Eq.~(\ref{mag}),
i.e., the dilaton is stabilized $\Phi=0$ and
the single component of ${\cal B}_{ij}$ is nonzero:
\begin{equation}\label{anb}
{\cal B}_{0i}=0,\quad {\cal B}_{23}(t)={\cal B}_{31}(t)=0,
\quad {\cal B}_{12}(t)\equiv B(t)\ne 0.
\end{equation}
The metric consistent with the matter, Eq.~(\ref{anb}), should keep
isotropy on the ($x^1-x^2$) plane and, thus, is of Bianchi type I:
\begin{equation}\label{me}
ds^2 = -dt^2 + a_1(t)^2 [(dx^1)^2 + (dx^2)^2] + a_3(t)^2(dx^3)^2.
\end{equation}
The energy-momentum tensor $T_{\mu\nu}$ of the
antisymmetric tensor field, Eq.~(\ref{anb}), is written in the form
\begin{eqnarray} \label{TB}
{(T_B)_\mu}^\nu &=& {(T_{\rm bulk})_\mu}^\nu + {(T_{\rm brane})_\mu}^\nu \\
&=& ({\tilde \Lambda}/{\tilde \kappa}_4^{2}){\delta_\mu}^\nu + {\rm
diag}\left[-\rho_B,-\rho_B,-\rho_B,+\rho_B\right] \nonumber\\
{}&+& {\rm
diag}\left[-\rho_b,-\tilde\rho_b,-\tilde\rho_b,-\rho_b\right],
\end{eqnarray}
where ${\tilde \Lambda}=m_{B}^{2}\Lambda$, ${\tilde \kappa}_{4}^{2}
=\kappa_{4}^{2}/m_{B}^{2}$, $\rho_B=\dot B^2/4{\tilde \kappa}_4^2a_1^4$,
$\rho_b=(1/2{\tilde \kappa}_4^2)\sqrt{1+B^2/a_1^4} \,$, and
$\tilde\rho_b=(1/2{\tilde \kappa}_4^2)/\sqrt{1+B^2/a_1^4}\,$.
Substituting Eqs.~(\ref{anb}), (\ref{me}), and (\ref{TB}) into
the Einstein equations, Eq.~(\ref{E-eq0}) or (\ref{Ee}) for the
tensor-field-dominated case are
\begin{eqnarray}
\left(\adoa1\right)^{2}+2\adoa1\adoa3 &=& {\tilde \kappa}_4^2(\rho_B+
\rho_b)+{\tilde \Lambda},
\label{ei1}\\
\addoa3+\addoa1+\adoa3\adoa1 &=& {\tilde \kappa}_4^2(\rho_B+\tilde\rho_b)
+{\tilde \Lambda}, \\
2\addoa1+\left(\adoa1\right)^{2} &=&
-{\tilde \kappa}_4^2(\rho_B-\rho_b)+{\tilde \Lambda}. \label{ei4}
\end{eqnarray}
The equation of motion for $B(t)$, Eq.~(\ref{B-eq0}) or (\ref{Be}), is
\begin{equation}\label{Beq}
\ddot B + \left(2\frac{\dot a_1}{a_1}+\frac{\dot a_3}{a_3}\right)\dot B
+\frac{B}{\sqrt{1+B^2/a_1^4}}=0.
\end{equation}

Introducing $\alpha_i=\ln a_i$, we can rewrite the Einstein equations,
Eqs.~(\ref{ei1})--(\ref{ei4}), as
\begin{eqnarray}
\dot\alpha_1^2+2\dot\alpha_3\dot\alpha_1
&=& {\tilde \kappa}_4^2(\rho_B+\rho_b)+{\tilde \Lambda},
\label{00}\\
\ddot\alpha_1+2\dot\alpha_1^2 +\dot\alpha_1\dot\alpha_3
&=& {\tilde \kappa}_4^2\rho_b+{\tilde \Lambda},
\label{11}\\
\ddot\alpha_3+\dot\alpha_3^{2}+2\dot\alpha_1\dot\alpha_3
&=& {\tilde \kappa}_4^2(2\rho_B+\tilde\rho_b) +{\tilde \Lambda}.
\label{33}
\end{eqnarray}
It is also convenient to introduce a new magnetic variable $b\equiv B/a_{1}^2$
by which we can rewrite the equation of the magnetic field, Eq.~(\ref{Beq}), as
\begin{equation}\label{Beq-2}
\ddot b+\left(2\dot\alpha_1+\dot\alpha_3\right)\dot b+
(2\ddot\alpha_1+2\dot\alpha_1\dot\alpha_3)b+\frac{b}{\sqrt{1+b^2}}
=0.
\end{equation}
Now let us examine  Eqs.~(\ref{00})--(\ref{Beq-2})
with and without the D3-brane and find cosmological solutions
under appropriate initial conditions of $b$.

\subsection{Anisotropy in the Bulk ($m_B=0$)}

In the absence of the D3-brane or in the limit of vanishing string coupling
$g_{{\rm s}}$, we recapitulate possible cosmological solutions.
This limit corresponds to the massless limit $m_B=0$ of the
antisymmetric tensor field so that the
derivative terms with the rescaled time variable ${\tilde t}=m_{B}t\rightarrow 0$
dominate in  Eqs.~(\ref{Beq})--(\ref{33}).
In this limit, the $B$-equation, Eq.~(\ref{Beq}), before the rescaling is
easily integrated to yield a constant of the motion,
\begin{equation} \label{L3}
\frac{a_3\dot B}{a_1^2} \equiv L_3 \ (=\textrm{constant}).
\end{equation}
With vanishing potential, $B$ manifests itself by
nonvanishing time derivatives. In the dual variable, it
corresponds to the homogeneous gradient along the $x^3$-direction. The
spacetime evolution with the dilaton rolling in this case was
studied in Ref.~\cite{copeland:1995}. Here, we have assumed that
the dilaton is stabilized by some mechanism.  Following
Ref.~\cite{copeland:1995} we introduce a new time coordinate
$\lambda$ via the relation $d\lambda=L_3dt/a_1^{2}a_3$.
Then, Eqs.~(\ref{00})--(\ref{33}) can be written as
\begin{eqnarray}
\label{Eeq123}
\alpha_1'\alpha_2'+\alpha_2'\alpha_3'+\alpha_3'\alpha_1'
&=& \frac14a_1^4, \\
\label{Eeq223}
\alpha_1''=\alpha_2'' &=& 0, \\
\label{Eeq333}
\alpha_3'' &=& \frac12a_1^4,
\end{eqnarray}
where the prime denotes the differentiation with respect to $\lambda$.

The solution for $\alpha_1$ is trivial; from Eq.~(\ref{Eeq223}),
\begin{equation}
\alpha_1 = C_1\lambda,
\end{equation}
where $C_{1}$ is constant, and we have omitted the integration
constant corresponding simply to rescaling of the scale factor. The
$\alpha_3$-equation, Eq.~(\ref{Eeq333}), is also easily integrated to
give
\begin{equation}
\alpha_3 = \frac{e^{4C_1\lambda}}{32C_1^2}+C_3\lambda .
\end{equation}
The constraint equation, Eq.~(\ref{Eeq123}), relates $C_{1}$ and $C_3$ by
$C_3=-C_1/2$.
Then, the relation between $\lambda$ and $L_3t$ is explicitly given by
\begin{eqnarray}
L_3t &=& \int^\lambda d\lambda\; a_1(\lambda)^{2}a_3(\lambda) \nonumber\\
&=& \int^\lambda d\lambda \exp\left[
(2C_1+C_3)\lambda+\frac{e^{4C_1\lambda}}{32C_1^2}\right]
\nonumber\\
&=& \left(16C_1\right)^{\frac{2C_1+C_3}{2C_1}}
\int^x dy\; y^{-P}e^y,
\end{eqnarray}
where $x=e^{4C_1\lambda}/32C_1^2$ and $P=(2C_1-C_3)/4C_1$.
The evolution of the scale factors for large $L_3t$ is given by
\begin{equation}
a_{1} \propto \left(\log L_3t\right)^{q_{1}}, \qquad
a_3 \propto L_3t,
\end{equation}
where $q_{1}=1/4$. Therefore, with non-vanishing
$B_{12}=B(t)$, only $a_3$ grows significantly, and a spatial
anisotropy develops. This can be seen clearly by considering the
ratio between Hubble parameters $H_1$ and $H_3$, $H_{3}/H_{1}$, where
$H_i\equiv \dot{a}_i/a_i$
\begin{equation}
{H_{3}\over H_{1}}= {4 \log L_3 t},
\end{equation}
which grows as time elapses. Finally, we remark that such
anisotropy cannot be overcome by some other type of isotropic
energy density in an expanding universe.  If one assumes the isotropic
universe ($a_i=a$) is driven by, for example, the radiation energy density
$\rho_{{\rm R}}$,  then one finds $\rho_B \propto 1/a^2$ from
Eqs.~(\ref{TB}) and (\ref{L3}); thus, $\rho_B/\rho_{{\rm R}} \propto
a^2$, which implies that the late-time isotropic solution can be
realized only in a contracting universe \cite{copeland:1995}.

\subsection{Isotropy on the D3-brane}

Let us now take into consideration the effect of a space-filling
D-brane. To get sensible solutions, we fine-tune the bulk
cosmological constant term to cancel the brane tension; that is,
$\Lambda=-m_B^2/2$, so that the effective four-dimensional
cosmological constant vanishes.

If we read the $B$-matter part explicitly,
then the full Einstein equations, Eqs.~(\ref{00})--(\ref{33}),
are rewritten as follows:
\begin{eqnarray}
\dot\alpha_1^2+2\dot\alpha_1\dot\alpha_3 &=&
\frac14\left(\dot b+2\dot\alpha_1b\right)^2\nonumber\\
{}&+& \frac{1}{2}\left(\sqrt{1+b^2}-1\right),\label{Eeq-11} \\
\ddot\alpha_1+\dot\alpha_1\left(2\dot\alpha_1+\dot\alpha_3\right) &=&
\frac{1}{2}\left(\sqrt{1+b^2}-1\right),\label{Eeq-1} \\
\ddot\alpha_3+\dot\alpha_3\left(2\dot\alpha_1+\dot\alpha_3\right)
&=& \frac12\left(\dot b+2\dot\alpha_1b\right)^2\nonumber\\
{}&+&\frac{1}{2}\left(\frac{1}{\sqrt{1+b^2}}-1\right).\label{Eeq-3}
\end{eqnarray}
With the help of Eq.~(\ref{Eeq-11}), Eq.~(\ref{Beq-2}) can be written as
\begin{eqnarray}
\label{Beq-3}
\ddot b &+&\left(2\dot\alpha_1 +\dot\alpha_3\right)\dot b\\
{}&+&
\left[-4\dot\alpha_1^2+\left(\sqrt{1+b^2}-1
+\frac{1}{\sqrt{1+b^2}}\right)\right]b=0,\nonumber
\end{eqnarray}
which is valid only for the $B$-dominated case, but is more
convenient for numerical analysis. Here, we have set
${\tilde \kappa}_4\equiv1$.

First, we examine the evolution of $b(t)$ and the scale factors
qualitatively. Suppose $b$ starts to roll from an
initial value $b_0$ while the universe is isotropic in the sense
that $\dot\alpha_{10}=\dot\alpha_{30}$. We assume initially
$a_{10}=a_{30}=1$ ($\alpha_{10}=\alpha_{30}=0$) and $\dot B_0=0$
so that $b_0=B_0$ and $\dot b_0=-2\dot\alpha_{10}B_0$. While $b$
is much larger than unity, the rapid expansion of $\alpha_1$ due
to the large potential proportional to $m_B^2b$ drives $b$, in
feedback, to drop very quickly to a small value of order one. Our
numerical analysis in Fig.~4 shows that this happens within
$m_Bt<2$ up to reasonably large value of $b_0$ for which the
numerical solution is working. The behavior of $b(t)$ after this
point is almost universal irrespective of the initial value $b_0$
if it is much larger than unity.

Once $b$ becomes smaller than unity, the quadratic mass term
dominates over the expansion, and $b$ begins to oscillate about
$b=0$. Then, the expansion of the universe provides a slow
decrease in the oscillation amplitude. The situation is the same
as that of a coherently oscillating scalar field, such as the
axion, or the moduli in the expanding universe. For small $b$, the
energy-momentum tensor of the oscillating $B$ field is given by
${T_\mu}^\nu={\rm diag}[-\rho,p_1,p_2,p_3]$, where
\begin{eqnarray}
\rho &=& \frac14\left(\dot b+2 \dot\alpha_1b\right)^2
+ \frac12\left(\sqrt{1+b^2}-1\right)\nonumber\\
{}&\approx & \frac14\left(\dot b^2+b^2\right), \\
p_1=p_2 &=& -\frac14\left(\dot b+2\dot\alpha_1b\right)^2
- \frac12\left(\frac{1}{\sqrt{1+b^2}}-1\right)\nonumber\\
{}&\approx & -\frac14\left(\dot b^2-b^2\right), \\
p_3 &=& \frac14\left(\dot b+2\dot\alpha_1b\right)^2
- \frac12\left(\sqrt{1+b^2}-1\right)\nonumber\\
{}&\approx & \frac14\left(\dot b^2-b^2\right).
\end{eqnarray}
With the expansion of the universe neglected, the equation of
motion for $b$ is then approximated by
\begin{equation}
\label{Beq-osc}
\ddot b+b\approx0.
\end{equation}
Since the oscillation is much faster than the expansion, we can
use the time-averaged quantities over one period of oscillation
for the evolution of spacetime. Equation (\ref{Beq-osc}) gives
the relation $\langle\dot b^2\rangle=\langle b^2\rangle$.
Thus, the oscillating $B$-field has the property
$p_1,p_2,p_3\approx0$ and behaves like homogeneous, isotropic
matter. This justifies the name of {\em B-matter}. Therefore,
after $b$ begins to oscillate, the isotropy of the universe is
recovered.

To quantify how the universe recovers isotropy, let us define the
parameter
$$ s \equiv \sqrt{2} \frac{H_1-H_3}{2 H_1+H_3} .$$
When $H_1\approx H_3$, the evolution of the quantity $s(t)$ is
determined by
\begin{equation}\label{ani}
\dot{s} = \frac16\left( \frac{4p_1-p_3-9 H^2}{H}\right) s +
\frac13\frac{p_1-p_3}{H} + {\cal O}(s^2),
\end{equation}
where $H\equiv \sum_i H_i/3$. At late time,
Eq.~(\ref{ani}) is approximated as $\dot{s} \approx -(3/2) H s$;
thus, one finds $ s\propto 1/t$.  Since $H$ is also proportional to
$1/t$, the initial and the final anisotropies $s_{i,f}$ follow the
relation
\begin{equation}
s_f = s_i \frac{H_f}{H_i},
\end{equation}
where $H_{i,f}$ denotes the initial and the final Hubble parameters,
respectively. To get an idea of how fast the anisotropy
disappears, let us consider $H_f \sim 10^{-15}$ GeV, which
corresponds to the Hubble parameter at the electroweak symmetry-breaking
scale of temperature, $T\sim 100$ GeV. Taking the initial
condition $s_i\sim 1$ around the beginning of the $B$ oscillation
$H_i \sim m_B$, we find that the final anisotropy can be
completely neglected for reasonable values of $m_B$.

For $m_B t\gg1$ with $H_1\simeq H_2 \simeq H_3$ and $b\ll1$, the
asymptotic solution of Eqs.~(\ref{Eeq-11})--(\ref{Beq-3}) can be
explicitly found to yield
\begin{equation}
H\propto 2/3t\,, \quad b \propto 1/t\,,\quad \rho_{B,b} \propto
1/t^2,
\end{equation}
which show the usual matter-like evolution. That is, the total
energy density of the $B$-matter diminishes like $\rho \propto
1/a^3$ even though the amplitude $B(t)$ itself grows like
$B(t)\propto a^{1/2}$.

Our qualitative results can be confirmed by  solving
Eqs.~(\ref{Eeq-1}), (\ref{Eeq-3}), and (\ref{Beq-3}) numerically.
Figure~4 shows the numerical solutions for the initial value
$b_0=100$. For a large value of $b_0$, $a_1$ grows very fast
while $a_3$ is frozen until $b$ becomes smaller than unity. Then,
$b$ begins to oscillate, and the universe becomes isotropic again
in that the expansion rates $\dot\alpha_1$ and $\dot\alpha_3$
converge; finally, the universe becomes $B$-matter dominated.

\begin{figure}[ht]
\begin{center}
\includegraphics[height=100pt]{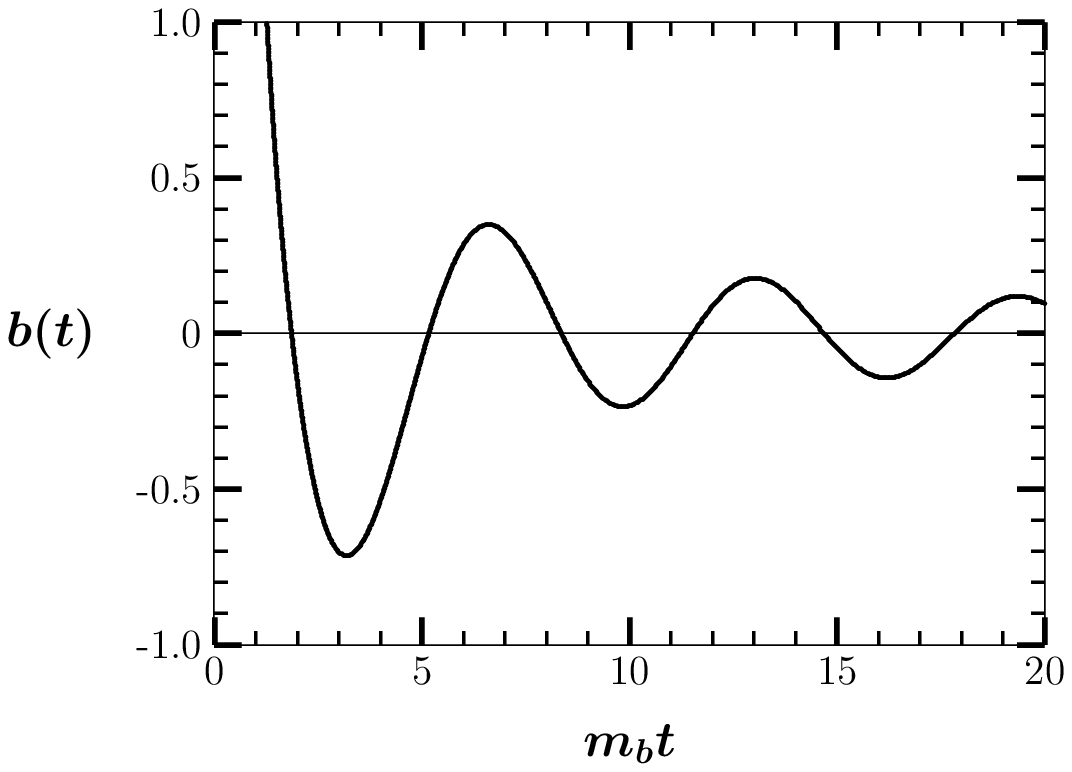}\\
\includegraphics[height=100pt]{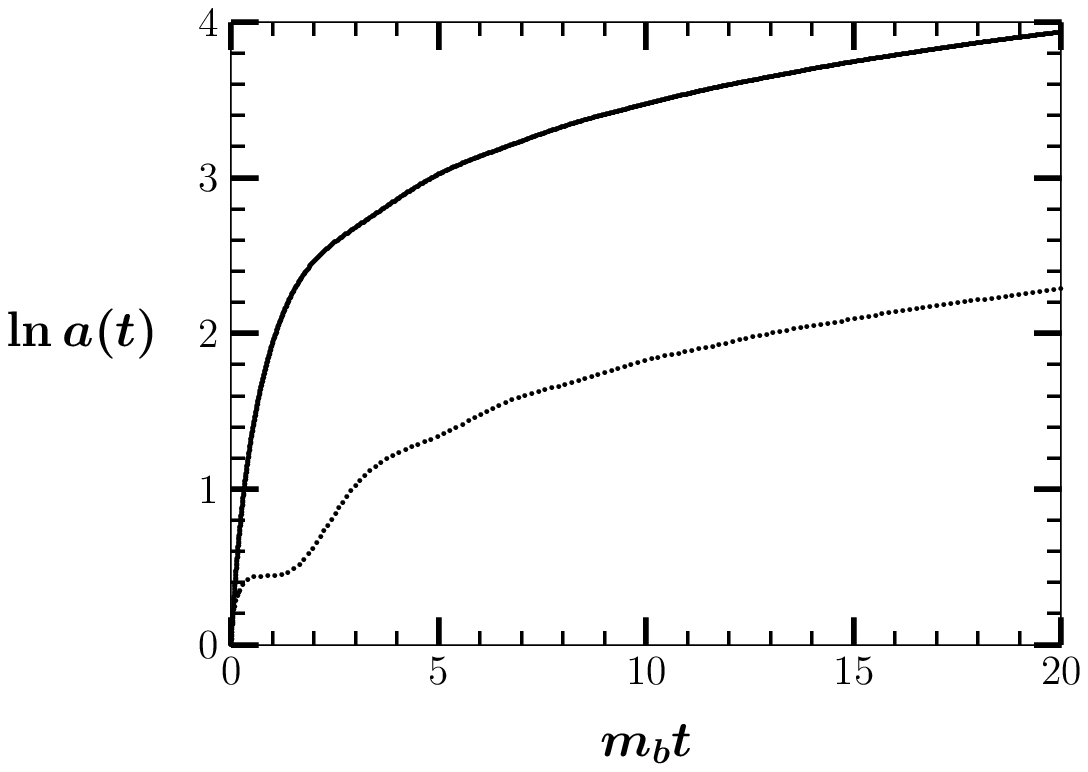}\\
\includegraphics[height=100pt]{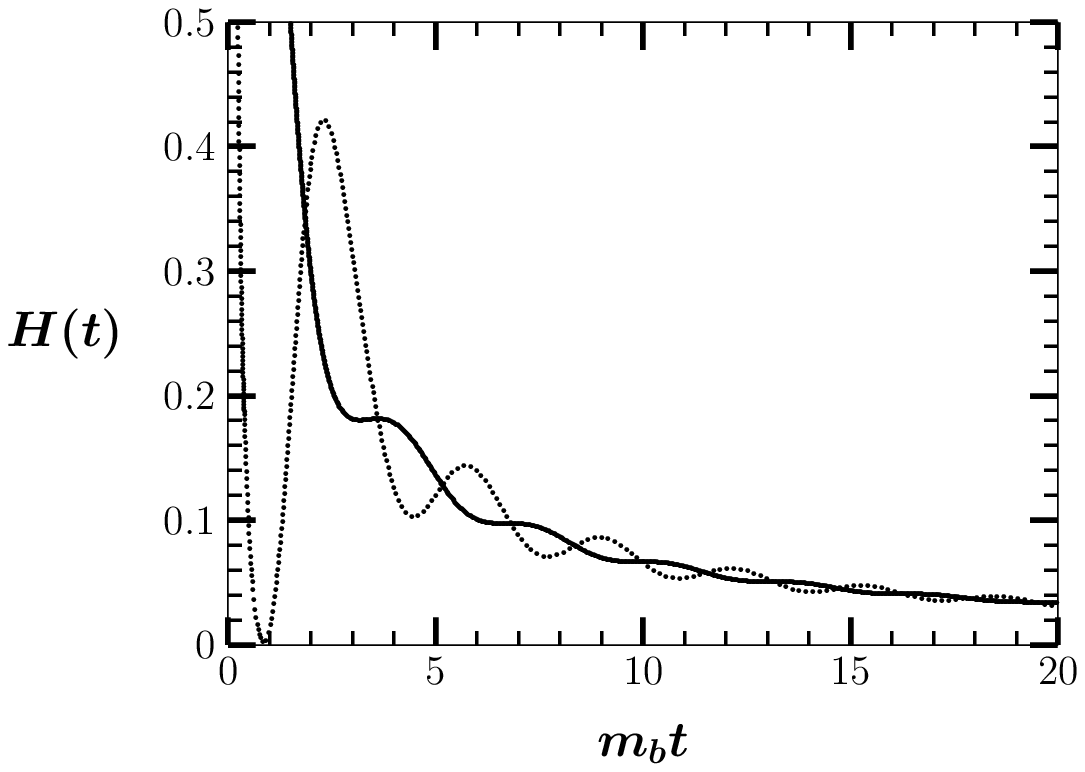}
\end{center}
\caption{Numerical solutions for $b_0=100$:
from the top down, the evolution of $b(t)$,
the scale factors $a_1(t)$ (solid line) and $a_3(t)$ (dotted line),
and the expansion rates $H_1(t)$ (solid line) and $H_3(t)$ (dotted line).}
\end{figure}

\section{Conclusion}
We considered the antisymmetric tensor field on a D3-brane. In a flat spacetime
with a stabilized dilaton, the condensed homogeneous magnetic field
oscillated without damping. When the linear dilaton was turned on,
we analyzed the difference between the string frame and the Einstein frame.
Particularly, the effect of the cosmological constant from (1+9)-dimensions
was drastic; i.e., the dilaton was stabilized for some initial values
and a negative cosmological constant~\cite{Kim:2005kr}.

In the early universe with gravitation, the existence of the D3-brane is
significant. In the bulk without the D-brane, the anisotropy induced by
the massless antisymmetric tensor field cannot be washed out through
the cosmological evolution,
but it remains as an observable fossil, which goes against
the observed fact in the present universe.
Once the mass and the self-interaction terms are included due to the presence
of D3-brane in our universe, even the huge condensation of the homogeneous
magnetic component is diluted as the universe expands. When the
cosmological expansion is sufficiently large,
the universe becomes isotropic and $B$-matter dominated.

It is intriguing to study cosmological evolution
when all the closed string degrees in the bulk
are included. Such a string cosmology
of the graviton, the dilaton, and the antisymmetric tensor field or that 
of unstable D-brane~\cite{Chung:2004ep} need further study.

\begin{acknowledgments}
This work was supported by the BK 21 project of the Ministry of
Education and Human Resources Development, Korea (I.C.),
the grant KRF-2002-070-C00022 (E.J.C.),
the research fund of Hanyang University (HY-2004-S) (H.B.K.),
and the Astrophysical Research Center for the
Structure and Evolution of the Cosmos (ARCSEC) supported by the KOSEF and
by Samsung Research Fund, Sungkyunkwan University, 2005 (Y.K.).
\end{acknowledgments}


\end{document}